\documentclass[amsmath,amssymb,prd,twocolumn]{revtex4}

\usepackage{graphicx}

\usepackage{citesort}
\include{aas_journals}

\newcommand{\be}{\begin{equation}}
\newcommand{\ee}{\end{equation}}
\newcommand{\bea}{\begin{eqnarray}}
\newcommand{\eea}{\end{eqnarray}}
\newcommand{\bean}{\begin{eqnarray*}}
\newcommand{\eean}{\end{eqnarray*}}

\def\yr{{\rm yr}}

\begin{document}
\title{Cosmic microwave background with Brans-Dicke gravity II:
constraints with the WMAP and SDSS data}
\author{Feng-Quan Wu }
\email{wufq@bao.ac.cn}
\author{Xuelei Chen}
\email{xuelei@cosmology.bao.ac.cn}

\affiliation{National Astronomical Observatories, Chinese
Academy of Sciences, \\
20A Datun Road, Chaoyang District, Beijing 100012, China}
\date{\today}
\begin{abstract}

Using the covariant formalism developed in a companion paper
\cite{wu:2008brans.theory}(paper I), we derive observational
constraint on the Brans-Dicke model in a flat
Friedmann-Lema\^{i}tre-Robertson-Walker (FLRW) universe with
cosmological constant and cold dark matter. The cosmic microwave
background (CMB) observations we use include the Wilkinson Microwave
Anisotropy Probe (WMAP) five year data, the Arcminute Cosmology
Bolometer Array Receiver (ACBAR) 2007 data, the Cosmic Background
Imager (CBI) polarization data, and the Balloon Observations Of
Millimetric Extragalactic Radiation and Geophysics (BOOMERanG) 2003
flight data. For the large scale structure (LSS) we use the matter
power spectrum data measured with the luminous red galaxy (LRG)
survey of the Sloan Digital Sky Survey (SDSS) Data Release 4 (DR4).
We parametrize the Brans-Dicke parameter $\omega$ with a new
parameter $\zeta=\ln(1/\omega+1)$, and use the Markov-Chain Monte
Carlo (MCMC) method to explore the parameter space. We find that
using CMB data alone, one could place some constraint on positive
$\zeta$ or $\omega$, but negative $\zeta$ or $\omega$ could not be
constrained effectively. However, with additional large scale
structure data, one could break the degeneracy at $\zeta<0$. The
$2\sigma$ (95.5\%) bound on $\zeta$ is $-0.00837<\zeta<0.01018$
(corresponding to $\omega < -120.0$ or $\omega > 97.8$). We also
obtained constraint on $\dot{G}/G$, the rate of change of $G$ at
present, as $-1.75 \times 10^{-12} \yr^{-1}<\dot{G}/G  < 1.05 \times
10^{-12}\yr^{-1}$, and $\delta G/G$, the total variation of $G$
since the epoch of recombination, as $ -0.083 < \delta{G}/G < 0.095$
at $2\sigma$ confidence level.

\end{abstract}


 \maketitle


\section{Introduction}

 The Jordan-Fierz-Brans-Dicke
theory\cite{Jordan:1949nature,Jordan:1959eg,Fierz:1956,Brans:1961sx,Dicke:1961gz}
(hereafter the Brans-Dicke theory for simplicity) is the most
natural alternative to the standard general relativity theory and
the simplest example of a scalar-tensor theory of
gravity\cite{Bergmann:1968ve,Nordtvedt:1970uv,Wagoner:1970vr,Bekenstein:1977rb,Bekenstein:1978zz}.
The gravitational constant becomes a function of space and time, and
is proportional to the inverse of a scalar field. Its action in the
usual (Jordan) frame is
\begin{equation}
\label{action} {\mathcal S}=\frac{1}{16\pi} \int d^4 x
\sqrt{-g}\left[-\phi R+
\frac{\omega}{\phi}g^{\mu\nu}\nabla_{\mu}\phi \nabla_{\nu}\phi
 \right] +{\mathcal S}^{(m)},
\end{equation}
where $\phi$ is the Brans-Dicke field,  $\omega$ is  a dimensionless
parameter,   and ${\mathcal S}^{(m)}$ is the action for the ordinary
matter fields ${\mathcal S}^{(m)}= \int d^4 x \sqrt{-g} {\mathcal L}^{(m)}.$
For convenience, we also define a dimensionless field
\begin{equation}
\varphi=G \phi,
\end{equation}
where $G$ is the Newtonian gravitational constant.
The Einstein equations are then generalized to
\begin{eqnarray} \label{generalized Einstein equations}
G_{\mu\nu} &=& \frac{8\pi G}{\varphi}T^{(m)}_{\mu\nu} +
\frac{\omega}{\varphi^2} (\nabla_\mu \varphi \nabla_\nu \varphi
-\frac{1}{2}g_{\mu\nu} \nabla_\lambda \varphi \nabla^\lambda \varphi
)
 \nonumber \\ &&
+\frac{1}{\varphi} (\nabla_\mu \nabla_\nu \varphi - g_{\mu\nu}
\nabla_\lambda \nabla^\lambda  \varphi),
\end{eqnarray}
where $T^{(m)}_{\mu\nu}$ is the stress tensor for all matter except
for the Brans-Dicke field, and the equation of motion for $\varphi$ is
\begin{equation}
\nabla_a\nabla^a \varphi= \frac{\kappa}{2\omega+3} T^{(m) \mu}_{\mu} ,
\label{eq:phievol}
\end{equation}
In order to match the
result of Cavendish type experiments, the present day value of
$\varphi$ should be
\bea \varphi_0=\frac{2\omega +4}{2\omega+3} . \label{phi 0}\eea

The original motivation of the Brans-Dicke theory is the idea that
the gravitational constant $G$ ought to be related to the average
value of a scalar field, which is determined by the mass density of
universe, so that the Mach principle is satisfied
\cite{Brans:1961sx,Dicke:1961gz}. Later, it is noted that the
scalar-tensor gravity appears in the low-energy limit of
supergravity theories from string theory
\cite{Green:superstringbook} and other higher-dimensional gravity
theories \cite{Appelquist:kaluzabook}. The Brans-Dicke field may be
associated with the dilaton-graviton sector of the string effective
action \cite{Green:superstringbook,Clarkson:2001qc}. The
dimensionless parameters in string theory - including the value of
the string coupling constant - can ultimately be traced back to the
vacuum expectation values of scalar fields
\cite{Becker:2006stringbook}.

The unexpected discovery of the accelerating expansion of the Universe
\cite{Riess:1998cb,Perlmutter:1997zf,Perlmutter:1998np}
forced us to look for an explanation of the so called ``dark energy'' which
may drive such acceleration.
Scalar fields rolling down a proper potential may serve as a
dynamical dark energy model \cite{Wetterich:1987fm,Peebles:1987ek,Frieman:1995pm,Turner:1998ex,Caldwell:1997ii,Liddle:1998xm,Steinhardt:1999nw}.
However, in these phenomenological models, the scalar fields are
added by hand, the connection to fundamental physics
is often unclear. The Brans-Dicke field $\phi$ is a natural candidate for the
scalar field, this is the so called ``extended quintessence''
scenario \cite{Uzan:1999ch,Amendola:1999qq,Chiba:1999wt,Perrotta:1999am,Holden:1999hm,Baccigalupi:2000je,Chen:2000xxa}. Alternatively,
the Brans-Dicke theory could
also serves as an effective model of the $f(R)$ gravity,
in which the gravity is invoked to explain the cosmic acceleration
\cite{Carroll:2003wy,Capozziello:2002rd,Capozziello:2003tk,Nojiri:2003ft,Nojiri:2003ni,Dolgov:2003px,Hu:2007nk,Dvali:2000hr,Deffayet:2001pu,Riazuelo:2001mg,EspositoFarese:2000ij,Bartolo:1999sq,Perrotta:1999am,Qiang:2004gg,Wu:2008rp}.

The Brans-Dicke theory is reduced to the Einstein theory in the limit of
\begin{equation}
\omega \to \infty, \quad  \varphi' \to 0,  \quad \varphi'' \to 0.
\end{equation}
So in some sense it could never be excluded completely if Einstein's
general relativity theory turns out to be the final words on the
classical theory of gravitation. So far, no significant deviation
from the Einstein theory has been discovered, and the most stringent
limit on the Brans-Dicke theory comes from solar-system experiments
which constrain the parametrized post-Newtonian (PPN) parameter
$\gamma=(1+\omega)/(2+\omega)$.  A recent significant result was
reported in 2003 using the Doppler tracking data of the Cassini
spacecraft while it was on its way to Saturn, with $\gamma-1=(2.1
\pm 2.3)\times 10^{-5} $ at  $2\sigma$ confidence level
\cite{Bertotti:2003rm}, which corresponds to  about $|\omega |>
40000$.  The limitation of such experiments is that they are
``weak-field" experiments and probe only a very limited range of
space and time. They could not reveal spatial or time variation of
gravitational constant on larger scales.

It has long been known that cosmological observations such as the
cosmic microwave background (CMB) and large scale structure (LSS) could
be used to test the Brans-Dicke
 theory \cite{Peebles:1970ag,Nariai:1969vh,Baptista:1996rr,Hwang:1996np,Liddle:1998ij,chenxuelei:1999brans,Nagata:2002tm,Nagata:2003qn,Acquaviva:2004ti,Acquaviva:2007mm,Schimd:2004nq,Tsujikawa:2008uc}. While the constraints obtainable
with such methods are generally weaker than the solar system tests, they probe
a much larger range of space and time. In recent years,
with the launch of the WMAP satellite, and the completion of the 2dF and
SDSS redshift surveys, it is interesting to put such test into practice.

In 2003, Nagata et al. used the WMAP first year data and $\chi^2$
test method to derive a constraint on the Brans-Dicke parameter.
They obtained $\omega>1000$ at $2\sigma$ confidence
level\cite{Nagata:2003qn}. However, in 2004, Acquaviva et al.
obtained a new constraint using a Markov Chain Monte Carlo approach
with CMB data from the WMAP first year data, the ACBAR, VSA and CBI
data, and the galaxy power spectrum data from 2dF. They obtain a
result of $\omega>120$ at $2\sigma$ confidence
level\cite{Acquaviva:2004ti}. These two limits differ by an order of
magnitude. We are unable to reproduce the result of
Ref.~\cite{Nagata:2003qn}, but we did reproduce successfully the
result of Ref.~\cite{Acquaviva:2004ti} using the procedures
described in their paper and the same data set as they used.

Nevertheless, as will be discussed in the next section, there is
room for improvement upon the method used in
Ref.~\cite{Acquaviva:2004ti}. Moreover, new CMB and LSS data have
since become available, it is therefore time to revisit this problem
with a new approach and update the constraint with the latest
observational data.

We have developed a covariant and gauge invariant method for calculating the
CMB anisotropy in Brans-Dicke theory, the formalism of our approach is
presented in the companion paper \cite{wu:2008brans.theory} (paper I).
In the present paper, we apply the
method developed in paper I, and use the latest CMB data and large scale
structure data to constrain the Brans-Dicke parameters. Here we consider
only the case of the massless Brans-Dicke model with
cold dark matter and cosmological constant. The more interesting
case of the Brans-Dicke field with interacting potential
would be investigated in future study.

\section{Methods}

The formalism of calculating CMB angular power spectra and matter power
spectrum in Brans-Dicke theory with the covariant and gauge invariant
method are presented in paper I. We also described in that paper the
numerical implementation of the method in the CMB code CAMB \cite{CAMB}.
The results of the modified CAMB code have been check with the results given
by Chen and Kamionkowski (1999) in Ref.\cite{chenxuelei:1999brans},
which was based on a modified version of CMBFAST in the synchronous
gauge. The output of the two code show excellent agreement. Our new code has been
implemented with some techniques to improve the architecture of program,
and the code is much faster than the old one. We refer the reader to
paper I for more details.

We consider deriving the constraint on the Brans-Dicke model with
the observational data using the Markov Chain Monte Carlo (MCMC) simulation.
The CAMB code is used by the public COSMOMC code \cite{Lewis:2002ah} as
the driver for calculating the CMB angular power spectra and matter power
spectrum. Here we use the modified CAMB code in the COSMOMC simulation.

The data we used to constrain the Brans-Dicke model are the latest
cosmic microwave background power spectrum data, which include the WMAP
five-year\cite{Nolta:2008ih}, ACBAR 2007\cite{Reichardt:2008ay}, CBI
polarization\cite{Sievers:2005gj} and BOOMERanG
2003\cite{Jones:2005yb,Piacentini:2005yq,Montroy:2005yx} data. We also
use the galaxy clustering power spectrum data derived from the SDSS LRG survey
DR4 \cite{Tegmark:2006az}.

We do not use the Type Ia supernovae (SNe Ia) data when
making the constraint in this paper,
because the value of the gravitational constant varies during the
expansion of the Universe. We know that the Chandrasekhar mass $M_{Ch}
\propto G^{-3/2}$. The variation of the gravitational constant $G$
means that the peak luminosity of SNe, which is approximately
pproportional to the Chandrasekhar mass, will change, so the supernovae
can not be assumed to be standard candles in this model.

Besides the Brans-Dicke parameter, the cosmological parameters
explored in our MCMC simulation are \{$\Omega_b h^2$, $\Omega_m
h^2$, $\theta$, $\tau$, $n_s$, $\log(10^{10} A_s), A_{SZ}$\}.  $
\Omega_b h^2$, $\Omega_m h^2$ are the baryon and matter densities
respectively. The $\theta$ parameter represents the ratio between
the sound horizon and the angular diameter distance to the last
scattering surface, it is used in lieu of the Hubble parameter $h$
since it is less correlated with other parameters. $\tau$ is the
optical depth to reionization, $A_s$ is the amplitude of primordial
superhorizon power spectrum in the curvature perturbation on $0.05
\mbox{Mpc}^{-1}$ scale, $n_s$ is the scalar spectral index, $A_{SZ}$
characterizes the
 marginalization factor of Sunyaev-Zel'dovich effect.
   We only consider the Brans-Dicke model in a flat universe
with the cosmological constant as dark energy. We assume flat priors
for these parameters, and the allowed ranges of the parameters are
wide enough such that further increasing the allowed ranges has no
impact on the results.

In any Bayesian approach to error estimate and parameter constraint,
the result will depend somewhat on the parametrization and prior.
The original Brans-Dicke parameter $\omega$ is inconvenient to use,
because it is unbounded, and the Einstein limit appears at
$\omega\to\infty$. Even if one restrict the allowed range of
$\omega$ to some finite interval, the large $\omega$ region would be
unduly favored, because at such region the difference in CMB and
LSS produced by models of different $\omega$ becomes indiscernibly
small.

Acquaviva et al. introduced a variable $\ln \xi= \ln [1/(4\omega$)]
in Ref.\cite{Acquaviva:2004ti}, and set its prior to be uniform in
the range $\ln\xi \in [-9, 3] $, corresponding to $ \omega \in[0.01,
2025.77] $.
The choice of the lower limit of $\ln \xi$ is motivated by the fact
that for $\omega>2000$, visual inspections show that the CMB angular
power spectra become insensitive to $\omega$. This parametrization
is workable, but has some drawbacks: firstly, it does not include
the negative values of $\omega$, and secondly, the lower limit of
$\ln \xi_{min} = -9 $, while ostensibly a reasonable choice, is
nonetheless put in by hand and quite arbitrary. In fact, the
$2\sigma$ limit would be sensitive to this artificial choice,
because the likelihood is high and almost flat at $\ln \xi<-9$, so
if one varies the lower limit $\ln\xi_{min}$, the overall
normalization of the posterior probability distribution function
would be directly affected.

In this paper, we introduce a new parameter which is more convenient to use:
\begin{equation}
\zeta=\ln(1+\frac{1}{\omega}).
\end{equation}
This parameter has the nice property that $\zeta \to 0$ asymptotes
the Einstein gravity, and it is easy to obtain the two-side (i.e.
allows negative $\omega$) likelihood distribution around $\zeta=0$.
$\zeta \approx 1/\omega $ when $\omega$ is a large number (i.e. very
close to Einstein gravity). We set the allowed range as $\zeta \in
[-0.014, 0.039]$, which brackets the Einstein gravity case, and
corresponding to $ \omega \in[-\infty, -71] \cup [25, \infty] $.
There is no arbitrary limit on large $|\omega|$ value, but only
limit on small $|\omega|$ value. Outside this range,  i.e.
$-71<\omega<25$, our numerical code break down, because the
background evolution deviates too much from the standard model.
However, as we are looking primarily for small departures from the
Einstein gravity, this is not a big concern, and large departures
would have been easily detected by other means as well. When making
plots of the likelihood, we do taken into account of the range of
allowed parameters, so that the probability is properly normalized.
Unavoidably, this artificial restriction on parameter range has some
effect on the final result, but as long as the final probability
distribution is much smaller than the allowed range, it would not
fundamentally change our conclusion.

\begin{figure}[htbp]
\begin{center}
\includegraphics[width=3.2in ]{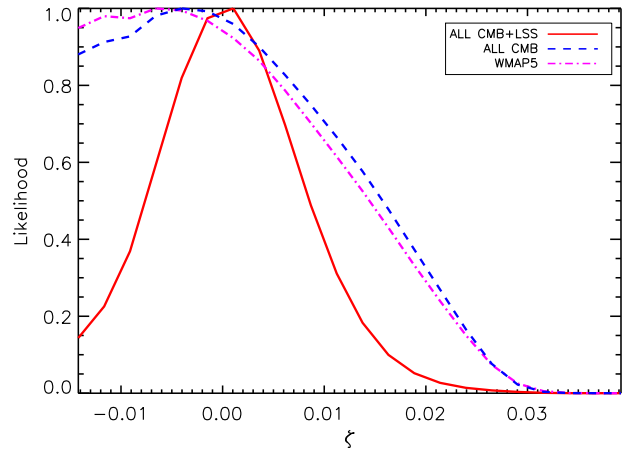}
\caption{The one dimensional marginalized  likelihood distributions
for the parameter $\zeta$. ``WMAP5" denotes WMAP five-year data.
``ALL CMB" represents WMAP five year data plus some small scale data
and polarization data, i.e. ACBAR 2007\cite{Reichardt:2008ay}, CBI
polarization\cite{Sievers:2005gj} and BOOMERanG
2003\cite{Jones:2005yb,Piacentini:2005yq,Montroy:2005yx} data.
``LSS" means galaxy clustering power spectrum data from SDSS DR4 LRG
data. } \label{fig:zeta 1D}
\end{center}
\end{figure}

\section{Results}
\subsection{Constraint on Brans-Dicke Theory}

The one-dimensional marginalized likelihood distributions for $\zeta$
is shown in Fig.\ref{fig:zeta 1D}. The three curves are obtained with
the WMAP data alone (magenta dash-dot curve), with all CMB data, i.e.
WMAP, ACBAR, CBI and Boomerang data (blue dashed curve),
and with all CMB data as well as the LSS data from SDSS LRG survey (red
solid curve). Interestingly, using only the CMB data, we find
that a negative $\zeta$ is favored. Indeed, the two
curves obtained with only the CMB data declines very slowly at $\zeta<0$,
making it difficult to obtain a limit on negative $\zeta$ with them, so
we could not easily quote a number for CMB-only constraint.
However, with the additional constraint
from the large scale structure data, the best fit value of
$\zeta$ goes back to the neighborhood of zero, and the likelihood
declines rapidly (almost Gaussian) at negative $\zeta$.
This shows that the large-scale structure data play an very
important role in constraining the Brans-Dicke gravity.

\begin{figure}[htbp]
\begin{center}
\includegraphics[width=3.2in]{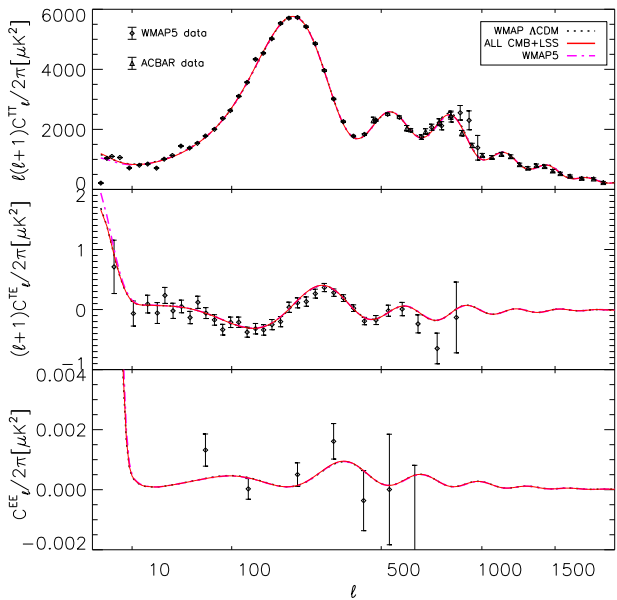}
\caption{CMB angular power spectra data with predictions of three
best fit models, see text for details. }
  \label{fig:spectrum.cmb}
\end{center}
\end{figure}
\begin{figure}[htbp]
\begin{center}
\includegraphics[width=3.2in]{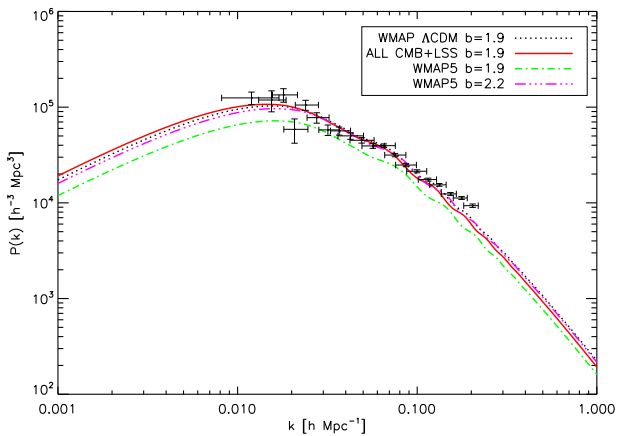}
\caption{The linear galaxy power spectra given by the three best-fit
models compared with SDSS LRG DR4 data\cite{Tegmark:2006az}.
We adopt original value of bias of $b=1.9$ as given in
Ref.\cite{Tegmark:2006az}. For the best fit model using only WMAP 5
year data, we also plot the result adjusting $b$ value to 2.2 to
better fit the galaxy power spectrum for comparison}
  \label{fig:spectrum.matt}
\end{center}
\end{figure}

To understand this result in more detail, we consider three models:

(1) the original best fit minimal 6-parameter $\Lambda$CDM model
with Einstein gravity obtained by the WMAP team using their 5 year
CMB data combined with the distance measurement from SN and the
Baryon Acoustic Oscillations(BAO) in the distribution of
galaxies\cite{Komatsu:2008hk}, which is marked as ``WMAP
$\Lambda$CDM" in the figure;
 (2) the best fit Brans-Dicke mode using only WMAP five-year CMB
 data, which is marked as ``WMAP5" in the figure;
 (3) the best fit Brans-Dicke model using all CMB data as well as
the SDSS LRG data, which  is marked as ``All CMB+LSS" in the figure.

The CMB angular
power spectra and linear galaxy power spectra for these models are
plotted in Fig.\ref{fig:spectrum.cmb} and
Fig.\ref{fig:spectrum.matt} respectively.

As shown in Fig.\ref{fig:spectrum.cmb}, due to parameter degeneracy,
the differences between the three curves of CMB are almost
indiscernible: for a slightly negative $\zeta$, the Brans-Dicke
model could produce CMB spectra which fits the data very well.
However, as shown in Fig.\ref{fig:spectrum.matt}, the matter power
spectra are quite different. The Brans-Dicke model which best-fit
the CMB data does not fit the galaxy power spectra very well. To be
sure, if one also allow the bias parameter as a free parameter, the
fit could be somewhat improved, nevertheless, it still fails
compared to the model obtained by fitting both the CMB and LSS data.
Thus, we see that the galaxy power spectrum data could play an
important role in distinguishing models, even though when used alone
its constraining power is relatively weak.

The 95\%  marginalized bound we derive in this paper is
\begin{equation}
 -0.00837<\zeta<0.01018
\end{equation}
corresponding to
\begin{equation}
 \omega < -120.0 ~~~~\rm{or} ~~~~ \omega > 97.8
\end{equation}
We note that when comparing this result with that of
Ref.~\cite{Acquaviva:2004ti}, one must remember that we have adopted
different parametrization and priors. In fact, we have used CMB data
with higher precision (WMAP 5 year vs WMAP 3 year), and additionally
we also used the LSS data (SDSS) which they did not use. Despite this
improvement in data quality, the limit we derived appears to be
slightly weaker than theirs, this is due to the different parametrization
and prior we used, particularly, we allowed negative $\omega$  which
was not considered in Ref.~\cite{Acquaviva:2004ti}.

\begin{figure}[htbp]
\begin{center}
\includegraphics[width=3.2in]{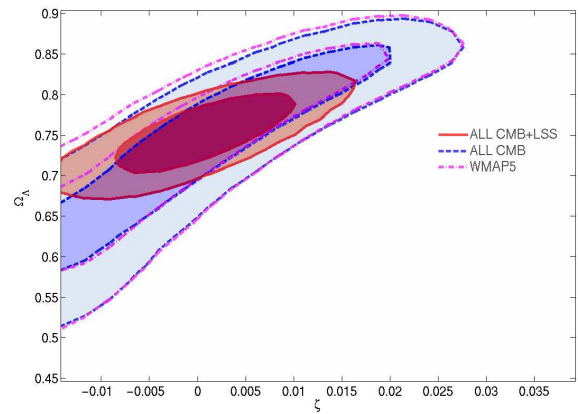}
\caption{The 2D contours of the marginalized likelihood distribution
of $\zeta$ against $\Omega_\Lambda$.  }
 \label{fig:zeta lambda 2D}
\end{center}
\end{figure}

To better understand the degeneracy and the 2-D likelihood
space distribution, we plot the 2-D contours of the marginalized
likelihood distributions of $\zeta$ against $\Omega_\Lambda$ in
Fig.\ref{fig:zeta lambda 2D}. The Einstein gravity with
$\Omega_{\Lambda}\sim 0.75$ is still the best fit model for the all CMB+LSS
data set. If $\zeta$ is greater,
$\Omega_{\Lambda}$ should also be greater, and vice versa.

\subsection{Constraint on cosmological parameters}

\begin{figure}[htb]
\centering{
\includegraphics[width=3.2in]{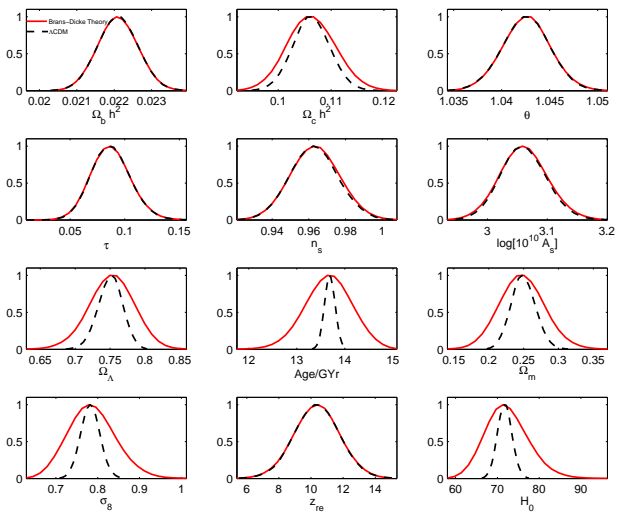}
\caption{ \label{fig:zeta multiply 1D} The one dimensional
marginalized probability distributions for the other cosmological
parameters in Brans-Dicke theory and in General Relativity. Data are
 all CMB data and the LSS data from SDSS
LRG survey(i.e. ALL CMB+LSS).   Red solid curves are results for
Brans-Dicke gravity, black dotted curves represent the case for
$\Lambda$CDM model in General relativity.
} }
\end{figure}

In Fig.\ref{fig:zeta multiply 1D}, we plot one-dimensional
marginalized likelihood distributions for other parameters in
Brans-Dicke theory(red solid curves), for comparison, we also plot
the same distributions in General Relativity case(black dotted
curves) which fixes $\xi=0$, using the same dataset---``ALL
CMB+LSS", i.e. all CMB data combined the LSS data from SDSS LRG
survey. The parameters in the top two rows of panels are the primary
cosmological parameters used in the MCMC program, and the parameters
in the bottom two rows of panels are the derived parameters (not the
parameters really run in the MCMC code). We see that the best fit
value of the parameters are almost unchanged. Furthermore, for most
of the primary parameters, the width of the likelihood distribution
is also unchanged. Only the distribution of the dark matter density
parameter $\Omega_c h^2 $ is slightly broader. For the derived
parameters, the best-fit values are also basically unchanged.
However, the likelihood distribution for most parameters are
broader, showing the introduction of the Brans-Dicke model allows
larger uncertainty in these parameters. The notable exception is the
reionization redshift $z_{re}$ which is basically unaffected.

\begin{figure*}[htbp]
\begin{center}
\includegraphics[width=6.4in]{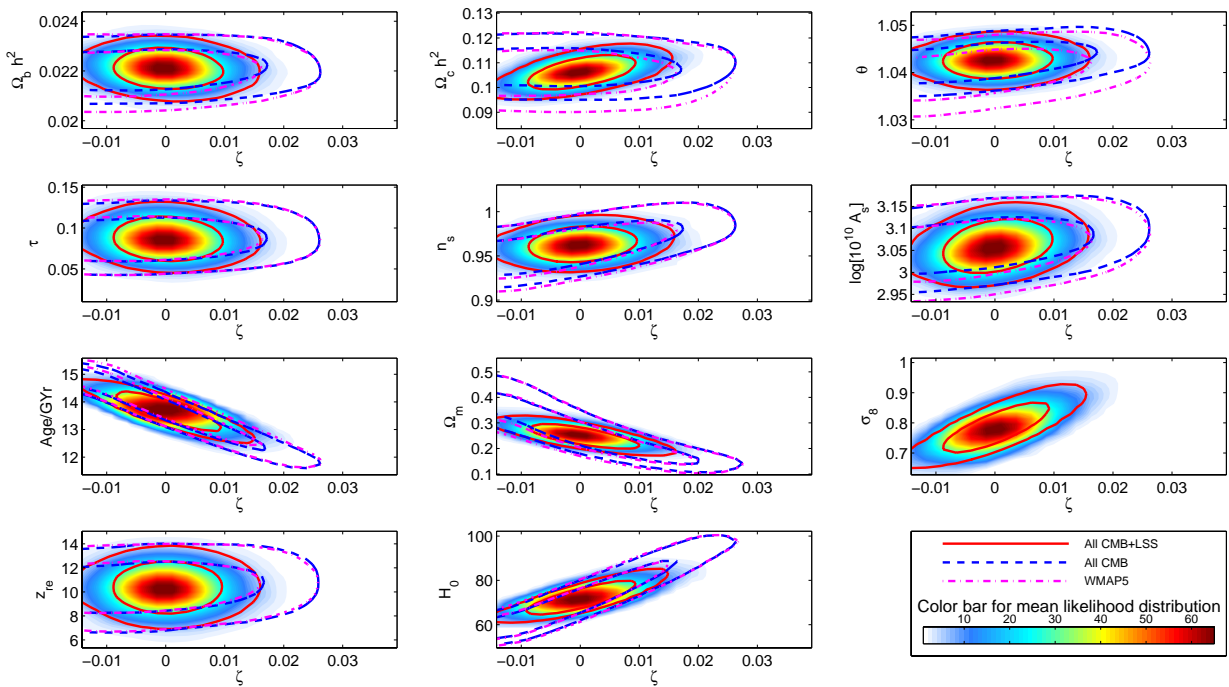}
\caption{ The 2D contours of the marginalized likelihood
distribution of $\zeta$ against the other cosmological parameters.
The color is for the mean likelihood distribution.}
 \label{fig:zeta multiply 2D}
\end{center}
\end{figure*}

The 2-D contours of of the marginalized likelihood distributions of
$\zeta$ against other cosmological parameters are shown in
Fig.\ref{fig:zeta multiply 2D}. As can be seen in the upper two rows
of panels, there are apparently not much correlations between
$\zeta$ and the other primary cosmological parameters used in MCMC
program, such as $\Omega_b h^2$, $\Omega_m h^2$, $\theta$, $\tau$,
$n_s$ and $\log(10^{10} A_s)$. However, from Fig.\ref{fig:zeta
lambda 2D} and the lower two rows of panels of Fig.\ref{fig:zeta
multiply 2D}, we see that $\zeta$ is correlated with
$\Omega_{\Lambda}$ and the derived parameters including the age of
Universe, $H_0$, $\Omega_m$, and $\sigma_8$, though there is almost
no correlation with the reionization redshift $z_{re}$.

We summarize the 68\% confidence limits on cosmological parameters
in Table \ref{tab:params}. Note that our pivot wavenumber $k_0=0.05
~\mbox{Mpc}^{-1}$ of the primordial power spectrum is different from
that of the WMAP group 5 year data release ($k_0=0.002~{\rm Mpc^{-1}}$),
and the set of primary parameters we used is also slightly different
from the one used by the WMAP group, as they used $\Omega_\Lambda$
instead of $\theta$ as a primary parameter. As we have mentioned,
the $\theta$ parameter is less correlated with $\zeta$, hence our
choice in this case could help improve the efficiency of the MCMC
method.  The data used by the WMAP group \cite{Komatsu:2008hk} are  the
WMAP five-year data, Type Ia supernovae data and the BAO data.
We have not included the supernovae data, which
we considered unreliable in the case of modified gravity.
From the Table \ref{tab:params}, we find that our best-fit
values of cosmological parameters are generally consistent with the
WMAP group result at one $\sigma$ confidence
level, however, our constraints are a bit weaker than those
given by the WMAP group, as we have added the Brans-Dicke parameter, and
also used somewhat different data sets.

\subsection{Constraint on the variation of gravitational constant $G$}

An interesting question is what limit could one place on
the variation of the gravitational constant $G$ using the CMB and
LSS observations. In the Brans-Dicke theory, $G$ also underwent
evolution from the time of recombination to the present time, the
variation in $G$ is correlated with the value of $\zeta$, so
we can also derive a limit on the variation of the $G$. Of course,
this evolution is not arbitrary,
but determined by the dynamical equation Eq.~(\ref{eq:phievol}), so
when citing the bounds on variation of $G$ obtained in this way, one
has to remember its limitations. Nevertheless, we note that in the Brans-Dicke
theory, the impact on CMB and LSS comes mainly from the variation
of $G$ \cite{chenxuelei:1999brans,wu:2008brans.theory}, so the result
obtained this way could still serve as a good reference value.

\begin{table*}
\begin{center}
\caption{\label{tab:params}Summary of cosmological parameters and
the corresponding 68\% intervals. }

{\scriptsize
\begin{tabular}{|c|c|c|c|c|c|c|c|}   \hline\hline

Class & Parameter & WMAP5 & ALL CMB & ALL CMB+LSS & WMAP group \cite{Komatsu:2008hk} \\
\hline

Primary & $\Omega_b h^2$ &$  0.02190_{ -0.00062}^{+  0.00073}$&$  0.02200_{ -0.00052}^{+  0.00069}$&$  0.02229_{ -0.00071}^{+  0.00033}$&$0.02265\pm 0.00059$ \\
& $\Omega_c h^2$&$   0.1040_{  -0.0049}^{+   0.0089}$&$   0.1064_{  -0.0039}^{+   0.0077}$&$   0.1066_{  -0.0046}^{+   0.0042}$& $0.1143\pm 0.0034$ \\
& $\theta$ &$   1.0391_{  -0.0024}^{+   0.0049}$&$   1.0425_{  -0.0028}^{+   0.0032}$&$   1.0432_{  -0.0030}^{+   0.0018}$&  \\
& $\tau$ &$    0.088_{   -0.009}^{+    0.009}$&$    0.085_{   -0.007}^{+    0.011}$&$    0.093_{   -0.014}^{+    0.001}$&$0.084\pm 0.016$ \\
& $n_s$ &$    0.947_{   -0.006}^{+    0.035}$&$    0.956_{   -0.012}^{+    0.027}$&$    0.962_{   -0.011}^{+    0.015}$&$ 0.960^{+ 0.014}_{- 0.013}$  \\
&$\log[10^{10} A_s]$&$    3.034_{   -0.024}^{+    0.074}$&$    3.050_{   -0.024}^{+    0.065}$&$    3.070_{   -0.047}^{+    0.030}$& $ A_s=(2.457^{+ 0.092}_{- 0.093})\times 10^{-9}$\\
\hline
Derived & $\Omega_\Lambda$&$     0.780_{    -0.009}^{+    0.100}$&$    0.789_{   -0.093}^{+    0.076}$&$    0.753_{   -0.031}^{+    0.029}$&$0.721\pm 0.015$ \\
&   $\Omega_b$ & & & & $0.0462\pm 0.0015$\\
&    $\Omega_c$ &&&&$0.233\pm 0.013$ \\
& Age/Gyr &$     14.09_{    -1.00}^{+  0.97}$&$    13.82_{    -1.13}^{+     0.82}$&$    13.63_{    -0.44}^{+     0.49}$&$13.73\pm 0.12\ \mbox{Gyr}$ \\
& $\Omega_m$&$     0.210_{   -0.085}^{+   0.085}$&$    0.218_{   -0.076}^{+    0.093}$&$    0.247_{   -0.029}^{+    0.031}$& \\
& $\sigma_8$ &&&$     0.789_{   -0.055}^{+    0.053}$&  $0.817\pm 0.026$ \\
&$z_{re}$&$  10.4_{  -1.5}^{+   1.7}$&$  10.2_{  -1.4}^{+   1.8}$&$  10.9_{  -1.8}^{+   0.9}$& $10.8\pm 1.4$\\
& $H_0$&$  63.5_{   -11.6}^{+  12.4}$&$  64.4_{  -9.7}^{+  14.2}$&$  72.3_{  -4.7}^{+   5.0}$& $70.1\pm 1.3$\\

\hline \hline
\end{tabular}
}
\end{center}
\end{table*}

\begin{figure}[htbp]
\begin{center}
\includegraphics[width=2.6in]{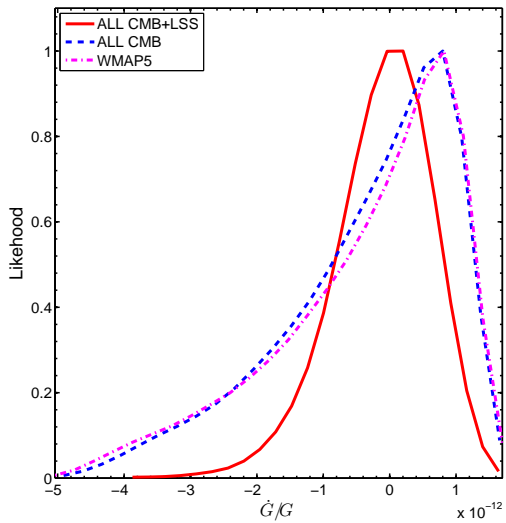}
\caption{One dimensional marginalized likelihood distributions of
$\dot{G}/G$.}
  \label{fig:dotg.1d}
\end{center}
\end{figure}

\begin{figure}[htbp]
\begin{center}
\includegraphics[width=2.6in]{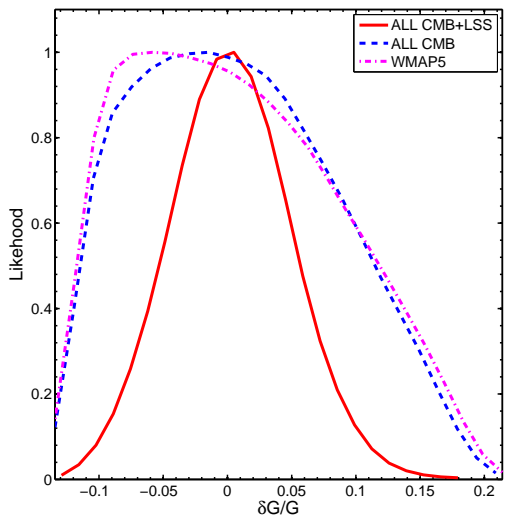}
\caption{One dimensional marginalized likelihood distributions of
$\delta{G}/G$.}
  \label{fig:deltag.1d}
\end{center}
\end{figure}

For making this constraint, we introduce two derived variables in
the MCMC, namely the rate of change of the gravitational constant
$\dot{G}/G$ at present and the integrated change of gravitational
constant since the epoch of recombination $\delta G/G$:
\begin{equation}
\dot{G}/G \equiv -\dot{\varphi}/\varphi, \qquad \delta G/G \equiv
(G_{rec}-G_0)/G_0.
\end{equation}
The one dimensional marginalized likelihood distributions of
$\dot{G}/G$ and $\delta{G}/G$ are plotted in Fig.~\ref{fig:dotg.1d}
and Fig.~\ref{fig:deltag.1d} respectively. The ``WMAP 5 year data''
and the ``all CMB data'' both favor a slightly non-zero (positive)
$\dot{G}/G$. With the addition of the SDSS power spectrum data,
however, the best-fit value is back to zero. From these figures, we
could still see some effect of the prior, as the likelihoods are
still non-zero or at best just approaching zero at the edge of the
figures. Nevertheless, with the LSS data added, the likelihood is
quite symmetric around the central value.

With this caveat in mind, we derive the following
$2\sigma$ (95.4\%) constraints:
\begin{equation}
-1.75 \times 10^{-12} \yr^{-1} < \dot{G}/G <  1.05 \times  10^{-12}\yr^{-1}
\end{equation}
and
\begin{equation}
-0.083 < \delta G/G <  0.095
\end{equation}

\begin{figure}[htbp]
\begin{center}
\includegraphics[width=3.2in]{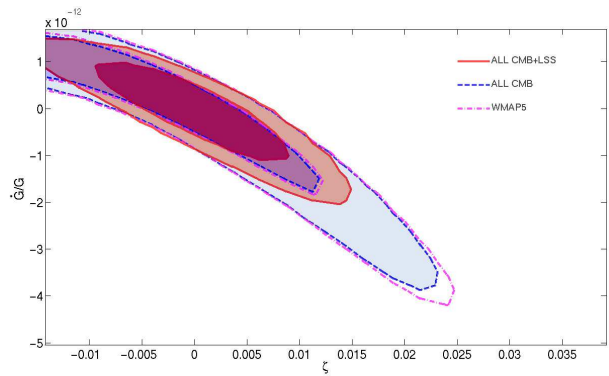}
\caption{2D contours of the marginalized likelihood distribution of
$\zeta$ against $\dot{G}/G$.}
  \label{fig:dotg.2d}
\end{center}
\end{figure}

\begin{figure}[htbp]
\begin{center}
\includegraphics[width=3.2in]{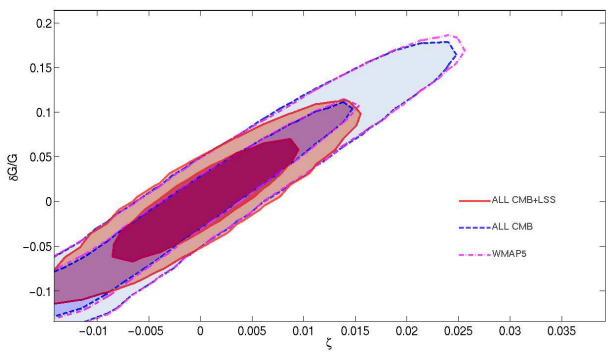}
\caption{2D contours of the marginalized likelihood distribution of
$\zeta$ against $\delta{G}/G$.}
  \label{fig:deltag.2d}
\end{center}
\end{figure}

We also plot 2D contours of  marginalized likelihood
distributions of $\zeta$ versus $\dot{G}/G$ and $\delta{G}/G$ in
Fig.~\ref{fig:dotg.2d} and Fig.~\ref{fig:deltag.2d} respectively.
As expected, the variation of gravitational constant is strongly
correlated with the value of $\zeta$ in this model.

Some previous constraints on these two variables together with the
result of the present paper are summarized in Table \ref{tab:G}.  We
note that in order to obtain such a constraint, one has to make some
assumptions, either in the underlying theoretical model, or in the
way $G$ varies. This is particularly true for the case of
constraints derived from CMB and LSS, as the impact of varying $G$
on these are multitude. For example, Ref.\cite{Chan:2007fe} modeled
the variations of $G$ by some hypothetical functions,
Ref.\cite{Galli:2009pr} parametrizes the evolution of $G$ as three
forms: constant, linear and Heaviside function,  while the present
paper assumed Brans-Dicke model. One has to be careful when
comparing the different limits, as the assumptions made are often
different. Nevertheless, from this table we can get a feeling of the
current limits on the variation of gravitational constants.

\begin{table*}
\begin{center}
\caption{\label{tab:G} Constraints on the rate of variations of
gravitational constant.
 The errors are $1\sigma$ unless otherwise noted. }

{\small
\begin{tabular}{|c|c|c|c|c|c|c|c|}   \hline\hline
Parameter & Value & Method & Reference \\
 \hline
$\dot{G}/G$ &$2\pm 7$ &lunar laser ranging& Muller \& Biskupek 2007\cite{Muller:2007zzb}\\
 \cline{2-4}
$[10^{-13} $yr$^{-1}]$ & $0\pm 4$ &big bang nucleosynthesis & Copi et al. 2004 \cite{Copi:2003xd}\\
                      &          &                         & Bambi et al. 2005\cite{Bambi:2005fi} \\
 \cline{2-4}
& $0\pm 16$  & helioseismology & Guenther et al. 1998 \cite{Guenther98} \\
 \cline{2-4}
&$-6\pm20$&neutron star mass& Thorsett 1996 \cite{Thorsett:1996fr} \\
 \cline{2-4}
&$20\pm40$ &Viking lander ranging & Hellings et al. 1983 \cite{hellings1983} \\
 \cline{2-4}
&$40 \pm 50$ & binary pulsar & Kaspi et al. 1994 \cite{Kaspi:1994hp} \\
 \cline{2-4}
&$-96\sim 81 ~(2\sigma)$& CMB (WMAP3)& Chang \& Chu 2007 \cite{Chan:2007fe} \\
 \cline{2-4}
&$-17.5\sim 10.5 ~(2\sigma)$& CMB+LSS & Wu \& Chen 2009 (This paper) \\
 \hline
 \hline

\end{tabular}
}
\end{center}
\end{table*}

\section{Conclusion}

In this paper, we use the currently available CMB (WMAP
five-year\cite{Nolta:2008ih}, ACBAR 2007\cite{Reichardt:2008ay}, CBI
polarization\cite{Sievers:2005gj} and BOOMERanG
2003\cite{Jones:2005yb,Piacentini:2005yq,Montroy:2005yx}) and the
LSS data (galaxy clustering power spectrum from SDSS DR4 LRG
data\cite{Tegmark:2006az}) to constrain the Brans-Dicke theory. We
use the covariant and gauge-invariant method developed in paper I to
calculate the CMB angular power spectrum and LSS matter power
spectrum.

To explore the parameter space, we use the MCMC technique. We
parametrize $\omega$ with a new parameter, $\zeta=\ln(1/\omega
+1)$, in order to explore the likelihood distribution of the
Brans-Dicke parameter $\omega$ in a continuous interval. This method
of parametrization is approximately equivalent to $\zeta=1/\omega$
when $\omega$ is a large  number.  It allows consideration of
negative $\omega $ value, and also there is no arbitrary upper limit
on $|\omega|$ (due to numerical problem, one has to choose a lower
limit for $|\omega|$). We explore in the range $ \zeta \in [-0.014,
0.039]$, corresponding to $ \omega \in[-\infty, -71] \cup [25,
\infty] $.

We found that while the CMB observation could constrain models with
positive $\omega$, {\it for the present data set and best fit
parameter values}, there is some degeneracy at $\omega<0$. The LSS
data could effectively remove this degeneracy. Finally, using the
CMB and LSS data, we obtain $2\sigma$ (95.5\%) limit on $\zeta$ as
$-0.00837<\zeta<0.01018$, corresponding to $\omega < -120.0$ or
$\omega > 97.8$. These limits may appear as weaker than previous
limit obtained by Ref.~\cite{Acquaviva:2004ti}, even though we used
later data than them. However, this difference is largely due to the
different assumption made in the constraint. Particularly, we
consider case of $\omega<0$ which was not considered in
Ref.~\cite{Acquaviva:2004ti}. As expected, the current limit on
$\omega$ derived from CMB and LSS data is much weaker than those
derived from solar system tests. However, the large temporal and
spatial range probed by these observations make it a useful
complementary to the latter.

To examine whether the gravitational coupling is really a constant
we introduced two new derived parameters in our MCMC code, one is
$\dot{G}/G$, the rate of change of the gravitational ``constant''
$G$ at present, and the other is $\delta G/G$, the integrated change
of $G$ since the epoch of recombination. We obtain the $2\sigma$
limit for these two variable as $-1.75 \times 10^{-12}
\yr^{-1}<\dot{G}/G < 1.05 \times  10^{-12}\yr^{-1}$ and $ -0.083 <
\delta{G}/G < 0.095$ respectively. These limits are still somewhat
weaker than the solar systems, but again they probed larger scales.
Especially for this test, the assumptions made in each technique
could be quite different, which one must bear in mind when making
comparisons.

The Planck satellite \footnote{{\it http://www.rssd.esa.int/index.php?project=planck}} is expect to begin operation and bring back
even better CMB data. The SDSS-3 BOSS
survey \footnote{{\it http://cosmology.lbl.gov/BOSS/, http://www.sdss3.org/ }},
WiggleZ \cite{Glazebrook:2007pp}, and the LAMOST surveys \cite{Wang:2008hgb}
are expected
to measure galaxy power spectrum at higher redshift and with better precision.
We look forward to obtain more stringent constraints on the Brans-Dicke
theory and other scalar-tensor gravity models in the near future.

\section*{Acknowledgements}

We thank  Antony Lewis, Le Zhang, Yan Gong, Xin Wang, Li'e Qiang,
G.F.R. Ellis and Marc Kamionkowski for helpful discussions. X.C.
acknowledges the hospitality of the Moore center of theoretical
cosmology and physics at Caltech, where part of this research is
performed. Our MCMC chain computation was performed on the
Supercomputing Center of the Chinese Academy of Sciences and the
Shanghai Supercomputing Center. This work is supported by the
National Science Foundation of China under the Distinguished Young
Scholar Grant 10525314, the Key Project Grant 10533010; by the
Chinese Academy of Sciences under grant KJCX3-SYW-N2; and by the
Ministry of Science and Technology under the National Basic Science
program (project 973) grant 2007CB815401.

\bibliography{brans}
\bibliographystyle{apsrev}
\end{document}